\documentclass[preprint,prc,groupedaddress,amsfonts,10pt]{revtex4}
\usepackage{amssymb}
\usepackage{dsfont}
\usepackage{amsmath}

\renewcommand{\arraystretch}{0.8}

\DeclareMathOperator{\Tr}{Tr}

\begin{document}

\title{Quantum $SU(3)$ Skyrme model with noncanonical embedded $SO(3)$ soliton}
\author{D. Jur\v ciukonis}
\email[]{darius@itpa.lt}
\author{E.Norvai\v sas}
\email[]{norvaisas@itpa.lt} \affiliation{
Vilnius University Institute of Theoretical Physics and Astronomy,\\
Go\v stauto 12, Vilnius 01108, Lithuania}

\begin{abstract}
The new ansatz which is the $SO(3)$ group soliton was defined for
the $SU(3)$ Skyrme model. The model is considered in noncanonical
bases $SU(3)\supset SO(3)$ for the state vectors. A complete
canonical quantization of the model have been investigated in the
collective coordinate formalism for the fundamental $SU(3)$
representation of the unitary field. The independent quantum
variables manifold cover all the eight dimensions $SU(3)$ group
manifold due to the new ansatz. The explicit expressions of the
Lagrangian and Hamiltonian densities are derived for this modified
quantum skyrmion.
\end{abstract}

\maketitle

\section{Introduction}

The topological soliton models, and the Skyrme model \cite{Skyrme}
among them, have generated a rising interest. It is related to
expansive field of applications of the model. The traditional
phenomenological application the model both in nuclear and
elementary particle physics as well as the Skyrme model
description of the quantum Hall effect \cite{Sondhi},
Bose-Einstein condensate \cite{Khawaja} and black hole physics
\cite{Shiiki}, are elaborated.

The first comprehensive phenomenological application of the model
to nucleon structure was the semiclassical calculation of the
static properties of the nucleon in \cite{Adkins}. The original
model was defined for a unitary field $U(\mathbf{x,}t)$ that
belongs to fundamental representation of the $SU(2)$ group.
Semiclassical quantization suggests that the skyrmion rotates as a
"rigid body" and the mass of the pion (asymptotic behavior of the
mass density) are introduced through an external chiral symmetry
breaking term in the Lagrangian density. Constructive realization
of \textit{ab initio} quantization provides in Hamiltonian a term
which may be interpreted as an effective pion mass term
\cite{Fujii,Acus}. The quantum mass corrections stabilize solution
of quantum skyrmion. The model was generalized to unitary field in
arbitrary irreducible representation (irrep) of the $ SU(2)$ group
\cite{Acus} and the $SU(3)$ group \cite{Jurciukonis}. The
extension of the Skyrme model to $SU(N)$ group \cite{Walliser}
represents the common structure of the Skyrme Lagrangian. The
intriguing rich geometrical structure with polyhedral symmetry
\cite{Manton} for winding (baryon) number larger than $1$ gives
impetus to wide applications of the model.

The aim of this work is to discuss the group-theoretical aspects
of the canonical quantization of the $SU(3)$ Skyrme model with new
$SO(3)$ ansatz which differs from proposed by A.P.Balachandran
\textit{et al} \cite{Balachandran}. The ansatz is defined in
noncanonical $SU(3)\supset SO(3)$ bases. These bases were
developed by J. P. Elliott to consider collective motion of nuclei
\cite{Elliott}.

The present manuscript is organized as follows. In Section 2 we
introduce noncanonical $SU(3)\supset SO(3)$ bases and $SO(3)$
hedgehog ansatz. In Section 3 the $SO(3)$ soliton is canonically
quantized on $SU(3)$ manifold. The explicit expression of the
Lagrangian and Hamiltonian densities of the quantum skyrmion are
presented in Section 4. The results are discussed in Section 5.

\section{Definitions for soliton in noncanonical SU(3) bases}

The unitary field $U(\mathbf{x,}t)$ is defined for $SU(3)$ group
in the arbitrary irrep $(\lambda ,\mu ).$ The modified Skyrme
model is based on standard Lagrangian density
\begin{equation}
{\mathcal{L}}=-\frac{f_{\pi }^{2}}{4}\Tr\{\mathbf{R}_{\mu
}\mathbf{R}^{\mu
}\}+\frac{1}{32\mathrm{e}^{2}}\Tr\{[\mathbf{R}_{\mu
},\mathbf{R}_{\nu }][\mathbf{R}^{\mu },\mathbf{R}^{\nu }]\},
\label{a1a}
\end{equation}
where the "right" and "left" chiral currents are defined as
\begin{eqnarray}
R_{\mu } &=&\left( \partial _{\mu }U\right)
\overset{+}{U}=\partial _{\mu }\alpha ^{i}C_{i}^{(B)}(\alpha
)\left\langle \hspace{0.2cm}\left\vert
\hspace{0.1cm}J_{(B)}^{(1,1)}\hspace{0.05cm}\right\vert
\hspace{0.2cm}\right\rangle ,  \label{a1b} \\
L_{\mu } &=&\overset{+}{U}\left( \partial _{\mu }U\right)
=\partial _{\mu }\alpha ^{i}C_{i}^{\prime (B)}(\alpha
)\left\langle \hspace{0.2cm}\left\vert
\hspace{0.1cm}J_{(B)}^{(1,1)}\hspace{0.05cm}\right\vert
\hspace{0.2cm}\right\rangle   \label{a1c}
\end{eqnarray}
and have the values on the $SU(3)$ algebra \cite{Jurciukonis}. The
$f_{\pi }$ and $\mathrm{e}$ in (\ref{a1a}) are the
phenomenological parameters of the model. The explicit expressions
of functions $C_{i}^{(B)}(\alpha )$ and $C_{i}^{\prime (B)}(\alpha
)$ depend on fixing of eight parameters $\alpha ^{i}$ of the
group. $J_{(B)}^{(1,1)}$ are the generators of the group in the
irrep $(\lambda ,\mu )$. At this work we will consider the unitary
field $U(\mathbf{x},t)$ in fundamental representation $(1,0)$
based on modified ansatz.

The Skyrme model Lagrangian for the $SU(2)$ group have the same
structure as (\ref{a1a})-(\ref{a1c}) only the chiral currents are
defined on \ $SU(2)$ algebra. We suggest the generalization of
usual hedgehog ansatz for any irrep $j$ of $SU(2)$ group
\cite{Norvaisas}
\begin{equation}
\exp i(\mathbf{\sigma }\cdot \hat{x})F(r)\rightarrow \exp
i2(\hat{J}\cdot \hat{x})F(r)=U_{0}\left( \hat{x},F(r)\right)
=D^{j}(\hat{x},F(r)), \label{a1}
\end{equation}
here $\hat{x}$ is the unit vector, $F(r)$ is a chiral angle
function, $\sigma $ - Pauli matrices, $\hat{J}$ - generators of
$SU(2)$ group in irrep $j$. The particular Wigner $D^{j}$ matrix
elements which represent the hedgehog ansatz for irrep $j$ are
\begin{equation}
D_{a,a^{\prime }}^{j}(\hat{x},F(r))=\frac{2\sqrt{\pi
}}{2j+1}w_{l}^{j}(F)\left[
\begin{array}{ccc}
j & l & j \\
a & m & a^{\prime }
\end{array}
\right] Y_{l,m}(\theta ,\varphi ),  \label{a2}
\end{equation}
were the symbol in brackets denotes the $SU(2)$ Clebsch - Gordan
coefficient, $Y_{l,m}(\theta ,\varphi )$ are the spherical
harmonics. The boundary conditions $F(0)=\pi$ and $F(\infty )=0$
ensures the winding (baryon) number $B=1$ for all irreps $j$ due
to the reason that the classical Lagrangian and winding number
have the same factor $N=\frac{2}{3}j(j+1)(2j+1)$ which can be
reduced \cite{Acus}. At this work we choose for the ansatz three
dimensional $SU(2)$ group representation which is the fundamental
$SO(3)$ group representation too. The radial dependent functions
in (\ref{a2}) for such ansatz are as follows:
\begin{eqnarray}
w_{0}^{1}(F) &=&\sqrt{2}\left( 3-4\sin ^{2}F\right), \notag\\
w_{1}^{1}(F) &=&i2\sqrt{3}\sin 2F,  \notag \\
w_{2}^{1}(F) &=&-4\sin ^{2}F.  \label{a3}
\end{eqnarray}

In this case it is convenient to define the noncanonical bases of
the $SU(3)$ irrep states vectors by reduction chain on subgroup
$SU(3)\supset SO(3)$. As we shall see later the structure of the
quantum skyrmion depends on a choice of bases for ansatz. For
general $SU(3)$ irreps $(\lambda ,\mu )$ the $SO(3)$ subgroup
parameters $(L,M)$ and its multiplicity can be sorted out by
different methods, see \cite{Tolstoy,Alisauskas}. Here considered
fundamental $(1,0)$ and adjoint $(1,1)$ representations of $SU(3)$
group are multiplicity free. The relations between canonical bases
(reduction chain $ SU(3)\supset SU(2)$) vectors $\left\vert
\begin{array}{c}
(1,0) \\
(z,j,m)
\end{array}
\right\rangle $ where hypercharge $y=\frac{2}{3}(\mu -\lambda )-2z$, and
noncanonical bases state vectors $\left\vert
\begin{array}{c}
(1,0) \\
(L,M)
\end{array}
\right\rangle $ are straightforward:
\begin{eqnarray}
\left\vert
\begin{array}{c}
(1,0) \\
(\frac{1}{2},\frac{1}{2},\frac{1}{2})
\end{array}
\right\rangle  &=&\left\vert
\begin{array}{c}
(1,0) \\
(1,1)
\end{array}
\right\rangle ;  \notag \\
\left\vert
\begin{array}{c}
(1,0) \\
(0,0,0)
\end{array}
\right\rangle  &=&\left\vert
\begin{array}{c}
(1,0) \\
(1,0)
\end{array}
\right\rangle ;  \notag \\
\left\vert
\begin{array}{c}
(1,0) \\
(\frac{1}{2},\frac{1}{2},-\frac{1}{2})
\end{array}
\right\rangle  &=&\left\vert
\begin{array}{c}
(1,0) \\
(1,-1)
\end{array}
\right\rangle .  \label{a4}
\end{eqnarray}
The system of noncanonical $SU(3)$ generator in terms of canonical
generators $J_{(Z,I,M)}^{(1,1)}$ which are defined in
\cite{Jurciukonis} can be expressed as follows:
\begin{eqnarray}
J_{(1,1)}=\sqrt{2}\left( J_{(\frac{1}{2},\frac{1}{2},\frac{1}{2}
)}^{(1,1)}-J_{(-\frac{1}{2},\frac{1}{2},\frac{1}{2})}^{(1,1)}\right)
\,,
&\qquad J_{(1,0)}=&2J_{(0,1,0)}^{(1,1)}\,, \notag \\
J_{(1,-1)}=\sqrt{2}\left(
J_{(-\frac{1}{2},\frac{1}{2},-\frac{1}{2}
)}^{(1,1)}+J_{(\frac{1}{2},\frac{1}{2},-\frac{1}{2})}^{(1,1)}\right)
\,,
&\qquad J_{(2,2)}=&-2J_{(0,1,1)}^{(1,1)}\,, \notag \\
J_{(2,1)}=-\sqrt{2}\left( J_{(\frac{1}{2},\frac{1}{2},\frac{1}{2}
)}^{(1,1)}+J_{(-\frac{1}{2},\frac{1}{2},\frac{1}{2})}^{(1,1)}\right)
\,,
&\qquad J_{(2,0)}=&-2J_{(0,0,0)}^{(1,1)}\,, \notag \\
J_{(2,-1)}=-\sqrt{2}\left(
J_{(-\frac{1}{2},\frac{1}{2},-\frac{1}{2}
)}^{(1,1)}-J_{(\frac{1}{2},\frac{1}{2},-\frac{1}{2})}^{(1,1)}\right)
\,, &\qquad J_{(2,-2)}=&2J_{(0,1,-1)}^{(1,1)}. \label{a5}
\end{eqnarray}
They satisfy the commutation relations
\begin{equation}
\left[ J_{(L^{\prime },M^{\prime })},J_{(L^{\prime \prime
},M^{\prime \prime })}\right] =-2\sqrt{3}\left[
\renewcommand{\arraystretch}{0.8}
\begin{array}{ccc}
\scriptstyle(1,1) & \scriptstyle(1,1) & \scriptstyle(1,1)_{a} \\
L^{\prime } & L^{\prime \prime } & L%
\end{array}%
\right] \left[
\begin{array}{ccc}
L^{\prime } & L^{\prime \prime } & L \\
M^{\prime } & M^{\prime \prime } & M%
\end{array}%
\right] J_{(L,M)}.  \label{6}
\end{equation}
The coefficients on the rhs of (\ref{6}) are $SU(3)$ noncanonical isofactor
and $SU(2)$ Clebsch - Gordan coefficient. The relations between the
noncanonical base state vectors and canonical state vectors for the adjoint
representation $(1,1)$ are similar to relations of generators (\ref{a5})
only with difference in normalization factor $\frac{1}{2}$ to keep the
states vectors normalized.

\section{Soliton quantization on SU(3) manifold}

We take the $SO(3)$ skyrmion (\ref{a2}) with $j=1$ as the
classical ground state $U_{0}$ for ansatz. The quantization of the
Skyrme model can be achieved by means of collective coordinates
$q^{\alpha }(t)$
\begin{equation}
U(\hat{x},F(r),q(t))=A(q(t))U_{0}\left( \hat{x},F(r)\right) A^{\dag }(q(t)).
\label{b1}
\end{equation}
We shall consider the Skyrme Lagrangian quantum mechanically
\textit{ab initio} \cite{Fujii} and eight unconstraint angles
$q^{\alpha }(t)$ to be quantum variables. Because the ansatz
$U_{0}$\ doesn't commute with all $SU(3)$ generators the
quantization is realized on eight parameter group manifold on the
contrary to the usual seven-dimensional homogeneous space
$SU(3)/U(1)$ \cite{Witten}. The generalized coordinates $q^{\beta
}(t)$ and velocities $(d/dt)q^{\alpha }\left( t\right)
=\dot{q}^{\alpha }\left( t\right)$ satisfy the commutation
relations
\begin{equation}
\left[ \dot{q}^{\alpha },q^{\beta }\right] =-if^{\alpha \beta }(q).
\label{b2}
\end{equation}
Here the symmetric tensor $f^{\alpha \beta }(q)$ is a function of
coordinates $q$ only. The explicit form of this tensor is
determined after the quantization condition has been imposed.

After substitution of the ansatz (\ref{a2}) into model Lagrangian
density (\ref{a1a}) it takes a form which is quadratic concerning
the velocities $\dot{q}^{\alpha }$
\begin{equation}
L=\int {\mathcal{L}d}^{3}x\thickapprox \frac{1}{2}\dot{q}^{\alpha
}\tilde{g}_{\alpha \beta }(q,F)\dot{q}^{\beta }+\left[
(\dot{q})^{0}-terms\right] , \label{b3}
\end{equation}
where the metric tensor
\begin{equation}
g_{\alpha \beta }(q,F)=C_{\alpha }^{\prime (L,M)}(q)E_{(L,M)(L^{\prime
},M^{\prime })}(F)C_{\beta }^{\prime (L^{\prime },M^{\prime })}(q),
\label{b4}
\end{equation}
and
\begin{equation}
E_{(L,M)(L^{\prime },M^{\prime })}(F)=-(-1)^{M}a_{L}(F)\delta _{L,L^{\prime
}}\delta _{M,-M^{\prime }}.  \label{b5}
\end{equation}
The soliton moments of inertia are given as integrals over dimensionless
variable $\tilde{r}=ef_{\pi }r$
\begin{subequations}
\begin{align}
a_{1}(F)=& \frac{1}{\mathrm{e}^{3}f_{\pi }}\frac{8\pi }{3}\int
d\tilde{r}\tilde{r}^{2}\sin ^{2}F\left[ 1+F^{\prime
2}+\frac{1}{\tilde{r}^{2}}\sin
^{2}F\right] ,  \label{b61} \\
a_{2}(F)=& \frac{1}{\mathrm{e}^{3}f_{\pi }}\frac{8\pi }{5}\int
d\tilde{r}\tilde{r}^{2}\Bigl[\sin ^{2}F\left( 3+2\cos 2F+\left(
9+8\cos 2F\right)
F^{\prime 2}\right)  \notag \\
& +\left( 9+4\cos 2F\right) \frac{\sin
^{2}F}{\tilde{r}^{2}}\Bigl].\label{b62}
\end{align}
\end{subequations}
The $a_{1}(F)$ coincides to the $SU(2)$ soliton momenta of
inertia, nevertheless the $a_{2}(F)$ differs from the second
soliton momenta of inertia in the usual $SU(3)\supset SU(2)$
ansatz case. The use of noncanonical $SU(3)$ bases lead to momenta
(\ref{b61}, \ref{b62}) which contrast with $SO(3)$ soliton momenta
of inertia defined in \cite{Balachandran}.

The canonical momenta are defined as
\begin{equation}
p_{\beta }=\frac{\partial L}{\partial \dot{q}^{\beta }}=\frac{1}{2}\left\{
\dot{q}^{\alpha },g_{\alpha \beta }\right\} .  \label{b7}
\end{equation}
They are conjugate to coordinates and satisfy the commutation
relations $\left[ p_{\beta },q^{\alpha }\right] =-i\delta _{\alpha
\beta }$. The curl brackets in (\ref{b7}) denotes the
anticommutator. The commutation relations and (\ref{b7}) fixe the
explicit expressions of the functions in (\ref{b2})
\begin{equation}
f^{\alpha \beta }(q)=\left( g_{\alpha \beta }(q)\right) ^{-1}.  \label{b8}
\end{equation}

The eight right transformation generators are defined as
\begin{equation}
\hat{J}_{(L,M)}=\frac{i}{2}\left\{ p_{\alpha },C_{(L,M)}^{\prime \alpha
}(q)\right\} =\frac{i}{2}\left\{ \dot{q}^{\beta },C_{\beta }^{\prime
(L^{\prime },M^{\prime })}(q)\right\} E_{(L^{\prime },M^{\prime })(L,M)}.
\label{b901}
\end{equation}
They satisfy the commutation relations (\ref{6}). Here the
functions $C_{(L,M)}^{\prime \alpha }(q)$ are dual to function
defined in (\ref{a1c})
\begin{subequations}
\begin{eqnarray}
\sum_{\alpha }C_{(L,M)}^{\prime \alpha }(q)C_{\alpha }^{\prime (L^{\prime
},M^{\prime })}(q) &=&\delta _{(L,M)(L^{\prime },M^{\prime }),}  \label{b902}
\\
\sum_{L,M}C_{(L,M)}^{\prime \alpha }(q)C_{\alpha^{\prime }
}^{\prime (L,M)}(q) &=&\delta _{\alpha \alpha ^{\prime }}.
\end{eqnarray}
\end{subequations}
The action of right transformation generators on the Wigner matrix
of $SU(3)$ irrep is well defined
\begin{eqnarray}
\left[ \hat{J}_{(L,M)},D_{(\alpha ,L^{\prime },M^{\prime })(\beta
,L^{\prime \prime },M^{\prime \prime })}^{(\lambda ,\mu
)}(q)\right]  &=&\left\langle (\lambda ,\mu )\alpha ,L^{\prime
},M^{\prime }\left\vert \hat{J}_{(L,M)}\right\vert (\lambda ,\mu
)\alpha _{0},L_{0},M_{0}\right\rangle
 \notag \\
&&\times D_{(\alpha _{0},L_{0},M_{0})(\beta ,L^{\prime \prime
},M^{\prime \prime })}^{(\lambda ,\mu )}(q). \label{b91}
\end{eqnarray}
The indices $\alpha$ and $\beta$ label the multiplets of $(L,M)$.
The substitution of the ansatz (\ref{b1}) into model Lagrangian
density (\ref{a1a}) and integration over spatial coordinates leads
to effective Lagrangian in the form
\begin{eqnarray}
L_{eff} &=&\frac{1}{2a_{2}(F)}(-1)^{M}\hat{J}_{(L,M)}\hat{J}_{(L,-M)}+\left(
\frac{1}{2a_{1}(F)}-\frac{1}{2a_{2}(F)}\right)   \notag \\
&&\times (-1)^{m}\left( \hat{J}_{(1,m)}\cdot \hat{J}_{(1,-m)}\right)
-M_{cl}-\Delta M_{1}-\Delta M_{2}-\Delta M_{3},  \label{b10}
\end{eqnarray}
were $M_{cl}$ is classical skyrmion mass, $\Delta M_{k}=\int
\mathrm{d}^{3}x\Delta \mathcal{M}_{k}(F)$ are quantum corrections
to the semiclassical skyrmion mass:
\begin{subequations}
\begin{align}
\begin{split}
\Delta {\mathcal{M}}_{1} = &-\frac{2\sin
^{2}F}{a_{1}^{2}(F)}\Bigl[f_{\pi }^{2}\left( 2-\cos 2F\right)
+\frac{3}{e^{2}}\left( F^{\prime 2}\left( 2+\cos 2F\right)
+\frac{\sin ^{2}F}{r^{2}}\right) \Bigl]\,;  \label{16}
\end{split}
\\
\begin{split}
\Delta {\mathcal{M}}_{2} = &-\frac{2\sin
^{2}F}{a_{2}^{2}(F)}\Bigl[f_{\pi }^{2}\left( 14+11\cos 2F\right)
+\frac{3}{e^{2}}\Bigl( F^{\prime 2}\left(
42+41\cos 2F\right) \\
& +\left( 25+12\cos 2F\right) \frac{\sin ^{2}F}{r^{2}}%
\Bigl) \Bigl]\,;  \hspace{5cm} \label{16a}
\end{split}
\\
\begin{split}
\Delta {\mathcal{M}}_{3} = &-\frac{4\sin ^{2}F}{a_{1}(F)\cdot
a_{2}(F)}\Bigl[f_{\pi }^{2}\left( 4+\cos 2F\right)
+\frac{3}{e^{2}}\Bigl( F^{\prime 2}\left( 6+5\cos 2F\right) \\
& +\left( 1-\cos 2F\right) \frac{\sin ^{2}F}{r^{2}} \Bigl)
\Bigl]\,. \hspace{5cm} \label{16b}
\end{split}
\end{align}
\end{subequations}
Two operators in (\ref{b10}) are quadratic Casimir operators of
$SU(3)$ and $SO(3)$ groups. A simple structure of operators
permits to write the eigenfunctions in the next section.

\section{Structure of the Hamiltonian density and the
symmetry breaking term}

To find explicit expression of Lagrangian and Hamiltonian density
of the quantum skyrmion we take into account the explicit
commutation relations (\ref{b2}) and (\ref{b8}). Some lengthy
manipulation with ansatz yields the expression of Lagrangian
density
\begin{equation}
{\mathcal{L}}_{SK}=\mathcal{K}-{\mathcal{M}}_{cl}-\Delta
{\mathcal{M}}_{1}-\Delta {\mathcal{M}}_{2}-\Delta
{\mathcal{M}}_{3},  \label{c1}
\end{equation}
were the kinetic (operator) part of the Lagrangian density is as follows:
\begin{eqnarray}
\mathcal{K}
&=&\frac{4}{a_{L}^{2}(F)}(-1)^{M}\hat{J}_{(L,M)}\hat{J}_{(L,M^{\prime
})}\Bigl\{\frac{f_{\pi }^{2}}{4}\left( \delta _{-M,M^{\prime
}}-D_{-M,M^{\prime }}^{L}(\hat{x},F(r))\right)   \notag \\
&&+\frac{3}{\mathrm{e}^{2}}\left( \delta _{-M,M^{\prime
}}-D_{-M,M^{\prime }}^{L}(\hat{x},F(r))\right) \Bigl\{\left(
F^{\prime 2}-\frac{1}{r^{2}}\sin
^{2}F\right)   \notag \\
&&\times \frac{2\sqrt{\pi
}(2L+1)\sqrt{\frac{1}{2}l+1}}{\sqrt{3}(5-2L)\sqrt{2l+1}}(-1)^{L+M+\frac{1}{2}l+1}\left\{
\begin{array}{ccc}
1 & 1 & l \\
L & L & L
\end{array}
\right\}   \notag \\
&&\times \left[
\begin{array}{ccc}
L & L & l \\
M & M^{\prime } & m
\end{array}
\right] Y_{l,m}(\theta ,\varphi )+\frac{1}{r^{2}}\sin
^{2}F\frac{1}{(5-2L)}\delta _{-M,M^{\prime
}}\Bigl\}\Bigl\}.\hspace{1cm}  \label{c2}
\end{eqnarray}

The coefficient in curl brackets are $6j$ coefficient of the
$SU(2)$ group. In (\ref{c2}) the Wigner $D_{M,M^{\prime
}}^{L}(\hat{x},F(r))$ matrices are used which in fact are hedgehog
anzatz for irrep $L=1,2$ in (\ref{a2}) . For representation $L=2$
the radial dependent functions are:
\begin{eqnarray}
w_{0}^{2} &=&\left( 5-20\sin ^{2}F+16\sin ^{4}F\right) , \notag \\
w_{1}^{2} &=&i\sqrt{2}\left( \sin 2F+2\sin 4F\right) , \notag \\
w_{2}^{2} &=&-\frac{2\sqrt{2\cdot 5}}{\sqrt{7}}\left( 7-8\sin ^{2}F\right)
\sin ^{2}F, \notag \\
w_{3}^{2} &=&-i4\sqrt{2}\sin ^{2}F\sin 2F, \notag \\
w_{4}^{2} &=&\frac{8\sqrt{2}}{\sqrt{7}}\sin ^{4}F. \label{c3}
\end{eqnarray}
The maximal spherical functions which appear in (\ref{c2}) are
$Y_{4,m}(\theta ,\varphi )$.

We define the momentum density as $\mathcal{P}_{\beta
}=\frac{\partial \mathcal{L}}{\partial \dot{q}^{\beta }}.$ The
kinetic energy density is define as
$2\mathcal{K}=\frac{1}{2}\left\{ \mathcal{P}_{\beta
},\dot{q}^{\beta }\right\} $. And the Skyrme model Hamiltonian
density takes the form
\begin{equation}
\mathcal{H}=\frac{1}{2}\left\{ \mathcal{P}_{\beta },\dot{q}^{\beta
}\right\}
-{\mathcal{L}}_{SK}=\mathcal{K}+{\mathcal{M}}_{cl}+\Delta
{\mathcal{M}}_{1}+\Delta {\mathcal{M}}_{2}+\Delta
{\mathcal{M}}_{3}. \label{c4}
\end{equation}
The operator (kinetic) part of Lagrangian (\ref{b10}) and kinetic
part of corresponding Hamiltonian depend on quadratic Casimir
operators of $SU(3)$ and $SO(3)$ groups which are constructed
using right transformation generators (\ref{b901}). The
eigenstates of the Hamiltonian $H=\int \mathrm{d}^{3}x
\mathcal{H}$ are
\begin{equation}
\left\vert
\begin{array}{c}
(\Lambda ,\Theta ) \\
\alpha ,S,N;\beta ,S^{\prime },N^{\prime }
\end{array}
\right\rangle =\sqrt{\dim (\Lambda ,\Theta )}D_{(\alpha ,S,N)(\beta
,S^{\prime },N^{\prime })}^{\ast (\Lambda ,\Theta )}(q)\left\vert
0\right\rangle ,  \label{c5}
\end{equation}
were complex conjugate Wigner matrix elements of the $(\Lambda
,\Theta )$ representation depends on eight quantum variables
$q^{\beta }$. The indices $\alpha$ and $\beta$ label the
multiplets of $SO(3)$ group. $\left\vert 0\right\rangle$ denotes
the vacuum state. Due to the structure of the density operator
(\ref{c2}) the noncanonical soliton mass distribution has a
complex but well defined tensorial structure which depends on
radial functions $F(r)$ and spherical harmonics $Y_{l,m}(\theta
,\varphi )$ of order $l=1,2,3,4$.

The mass or energy functional of (\ref{c5}) state is as follows
\begin{eqnarray}
M &=&\frac{2}{3a_{2}(F)}\left( \Lambda ^{2}+\Theta ^{2}+\Lambda \Theta
+3\Lambda +3\Theta \right)    \notag \\
&&+\left( \frac{1}{2a_{1}(F)}-\frac{1}{2a_{2}(F)}\right)
S(S+1)+M_{cl}+\Delta M_{1}+\Delta M_{2}+\Delta M_{3}. \label{c51}
\end{eqnarray}
In contrast to the positive impact of Casimir operators (quantum
rotation) to the classical mass $M_{cl}$ the quantum corrections
$\Delta M$ which appear from commutation relations are negative.

We take account of chiral symmetry breaking effects by introducing
the term
\begin{equation}
\mathcal{M}_{SB}=\frac{1}{4N}f_{\pi }m_{0}^{2}Tr(U+U^{\dagger }-2),
\label{c6}
\end{equation}
which takes an explicit form
\begin{equation}
\mathcal{M}_{SB}=\frac{1}{2}f_{\pi }m_{0}^{2}\sin ^{2}F.  \label{c7}
\end{equation}

In (\ref{c6}) we used the same normalization factor $N=4$ which is
defined for $SO(3)$ classical soliton $j=1$.

The direct calculation shows that the Wess-Zumino-Witen term is
equal to zero $L_{WZ}=0$ for noncanonical embedded $SO(3)$
soliton.

\section{Conclusion}

In this paper we have considered a new ansatz for Skyrme model
which is noncanonical embedded $SU(3)\supset SO(3)$ soliton. The
strict canonical quantization of the soliton leads to new
expressions of momenta of inertia and negative quantum corrections
$\Delta M.$ The quantum corrections which appear from commutation
relations compensate the effect of positive $SU(3)$ and $SO(3)$
"rotation" kinetic energy. The variation of quantum energy
functional (\ref{c51}) allow to find the stable solutions of
quantum skyrmions even without symmetry breaking term. The shape
of quantum skyrmion are not fixed like in semiclassical "rigid
body" case and infinite tower of solutions for the higher
representations $(\Lambda ,\Theta )$ is absent. It means that the
"fast rotation" destroys the quantum skyrmion. For details of
canonical $SU(2)$ skyrmion quantization see \cite{Acus}. The
unitary field $U(\mathbf{x},t)$ for $SU(3)$ Skyrme model can be
define in arbitrary irrep $(\lambda ,\mu )$. The $SU(2)$ $\left(
SO(3)\right) $ ansatzes can be constructed as reducible
representations of $SU(2)$ embedded into $SU(3)$ irrep $(\lambda
,\mu )$ in different ways. It can generates different types of
quantum skyrmions.

\begin{acknowledgments}

The authors would like to thank S. Ali\v{s}auskas for discussions
on $SU(3)$ noncanonical bases.

\end{acknowledgments}


\begin{thebibliography}{99}
\bibitem{Skyrme} T.H.R. Skyrme, Proc. Roy. Soc. \textbf{A260}, 127 (1961).

\bibitem{Sondhi} S.L. Sondhi, A. Karlhede, S.A. Kivelson, and E.H. Rezayi,
Phys. Rev. B \textbf{47} 16419 (1993).

\bibitem{Khawaja} U. A. Khawaja and H. Stoof, Nature (London) \textbf{411},
918 (2001).

\bibitem{Shiiki} N. Shiiki and N. Sawado, Phys. Rev. \textbf{D71}, 104031
(2005).

\bibitem{Adkins} G.S. Adkins, C.R. Nappi, E. Witten, Nucl. Phys. \textbf{B228%
}, 552 (1983).

\bibitem{Fujii} K. Fujii, A. Kobushkin, N. Toyota, Phys. Rev. Lett. \textbf{%
58}, 651 (1987), Phys. Rev. D \textbf{35}, 1896 (1987).

\bibitem{Acus} A. Acus, E. Norvai\v{s}as, D.O. Riska, Phys. Rev. C \textbf{57%
}, 2597 (1998).

\bibitem{Jurciukonis} D. Jur\v{c}iukonis, E. Norvai\v{s}as, D.O. Riska, J.
Math. Phys. \textbf{46}, 072103 (2005).

\bibitem{Walliser} H. Walliser, Nucl. Phys. A \textbf{548}, 649 (1992).

\bibitem{Manton} N. Manton and P. Sutcliffe, \textit{Topological Solitons} (Cambride
University Press, Cambrige, 2004).

\bibitem{Balachandran} A. P. Balachandran, F. Lizzi, V. G. J. Rodgers, Nucl.
Phys.  \textbf{B256}, 525 (1985).

\bibitem{Elliott} J. P. Elliott, Proc. Roy. Soc. \textbf{A245}, 128 (1958).

\bibitem{Norvaisas} E. Norvai\v{s}as, D.O. Riska, Phys. Scr. \textbf{50},
634 (1994).

\bibitem{Tolstoy} V.N. Tolstoy, Phys. Atom. Nucl. \textbf{69}, 1058 (2006).

\bibitem{Alisauskas} S. Ali\v{s}auskas, J. Phys. A: Math. Gen. \textbf{20},
1045 (1987).

\bibitem{Witten} E.Witten, Nucl. Phys. \textbf{B223}, 422 (1983); \textbf{%
B223}, 433 (1983).

\end{thebibliography}
\end{document}